# BUILDING FINANCIAL ACCURACY INTO SPREADSHEETS

Andrew Hawker
Department of Accounting & Finance, University of Birmingham, UK
Email: A.Hawker@bham.ac.uk. Fax: 0121414 6678.

**ABSTRACT**

*Students learning how to apply spreadsheets to accounting problems are not always well served by the built-in financial functions. Problems can arise because of differences between UK and US practice, through anomalies in the functions themselves, and because the promptings of Wizards' engender an attitude of filling in the blanks on the screen, and hoping for the best. Some examples of these problems are described, and suggestions are presented for ways of improving the situation. Principally, it is suggested that spreadsheet prompts and 'Help' screens should offer integrated guidance, covering some aspects of financial practice, as well as matters of spreadsheet technique.*

## 1. INTRODUCTION

When spreadsheeting first came into the accounting curriculum, students spent a lot of time learning to deal with some awkward and temperamental features of the software. Everyone looked forward to the day when it would become more user-friendly. Perhaps this was a little naive. There is no doubt that, with today's products and the dominance of Microsoft's style, students can get to grips with spreadsheets much faster than they used to. Unfortunately, this can also give them a false sense of confidence, as they discover and activate all the powerful built-in functions.

Over the past decade or so, considerable evidence has accumulated of the dangers of poorly-constructed spreadsheets, thanks to the efforts of Brown [1], Panko [2], Janvrin [3], and others. A number of useful guides to good practice and audit are now available [4,5]. Most of these publications suggest ways in which those who create spreadsheets should be more systematic and vigilant. This clearly has to be the first line of attack in promoting good practice. However, there are still some occasions when the products themselves could help to pre-empt errors.

This paper is concerned with one particular aspect of spread-sheeting (the use of financial functions) in one particular niche in the education system (accounting courses in a university). If students are going to move on to use the financial functions in a business environment, it is important that they know what they are doing. Their coursework includes practice in the use of functions for calculating present values, interest rates, depreciation, and so on. A recurring problem is that the technology works smoothly enough, but the functions do not behave exactly as might be expected.

The comments which follow are based on experiences with Microsoft Excel, up to and including Excel 97. This is not intended to suggest that these problems are by any means confined to Microsoft's product.

## 2. THE PROBLEM AREAS

The problem areas can be grouped under three headings:

### 2.1. Unexpected results

A number of arguments have raged about arithmetic in spreadsheets when large numbers of significant figures are involved, but this is rarely an issue for accountants. Sometimes, however, the accuracy turns out to be less than n-fight be expected. For example, the declining balance function in Excel calculates and uses a factor which is accurate to three decimal places. This means that if you are dealing with an asset worth millions of pounds, the final few figures of each depreciation figure will be suspect.

Depreciation is not a very exact art, and this is fine so long as the user converts the results to round sums. Furthermore, it is unlikely that those doing such calculations manually in the past ever bothered to work out the factor beyond a few decimal places. Spreadsheets, though, are expected to deliver precision. Anyone who adds up depreciation values over the life of the asset, and expects that the salvage value will always equate to the initial value minus the accumulated depreciation, could be in for a shock. (A quick way of verifying this is to add together the values provided in the example given on the Help screen for the **DB** function). They could be further perplexed if they used the optional "month" argument in this function. This makes rather primitive adjustments for an asset bought part way through the first accounting year. Users of this argument have to bear in mind that (a) it only makes sense if you are working in periods of one year, (b) you need to add an extra period at the end (not mentioned in the Help text) and (c) your results are even *less* likely to balance than when you leave out this optional argument altogether. All this is fairly difficult for a novice student to take on board.

### 2.2. Misleading or ineffective "help"

Efforts to make Help facilities more "Intelligent" have not always been well received. Unsolicited interventions, particularly by cartoon paper-clips, can be extremely irritating. However, there are many areas of Help text which remain unchanged from early Excel releases. Where changes have been made, they often have the feel of amendments to a software specification, rather than real attempts to guide and inform the behaviour of users. For example, an extra option has been added to the **ACCRINT** and related functions, to allow users to choose a "European 30/360" method of calculation. In Europe and the US, the same basic formula is used to calculate interest on a "30/360" basis, but there are some potentially crucial differences in the way months are treated if they are not actually 30 days long. Originally, the Excel function only offered the US method, as its default.

This is an example of a situation where it would be very useful if users could, if they wished, be offered more explanation of the financial as well as the technical aspects of the functions they are using. They could be advised, for example, on the types of investments for which each option is appropriate, and the key differences between the different methods of calculation. Such explanations are to be found on the web (for example at **bondchannel.bridge.com).** In theory, we could build some appropriate hyperlinks into the spreadsheet. However, since many of the issues are basic and unchanging, it would be far better if they were incorporated into the routine Help material, alongside the normal guidance on technique. When actually using the spreadsheet it is, after all, necessary to consider all these different aspects of the problem at the same time.

Interest rates can cause other problems, again resulting mainly from the Atlantic divide. Help screens routinely advise that annual interest rates can be converted to monthly ones simply by dividing by 12. This advice is not tempered by any reference to effective versus nominal interest rates. It accords with much American practice, but causes problems if, for example, students are invited to replicate a loan repayment table issued by a UK bank. This is because UK practice is to quote effective annual rates (based on compounded monthly interest). Divide the annual rate by 12, and the results you obtain will be too high.

Excel has the **effective** and **nominal** functions to help get round these problems, but makes no effort to steer users towards them at the critical moment. Furthermore, recent supplementary advice from Microsoft on using different compounding periods [6] completely ignores this as a potential pitfall.

This leads off into some other interesting territory, since UK lenders are permitted, by law, to quote annual rates rounded down to a single decimal point (so that many of them routinely set rates such as 11.995%, which can legitimately be advertised as 11.9%). It is quite a good test of students' audit skills to have them replicate the calculations, based on the advertised figures, and pick up on anomalies of this kind. In one particular loan scheme, for example, customers were advised that they could defer paying back the loan during a repayment "holiday". What the advertising omitted to mention was that interest on the full amount of the advance was charged during the "holiday" period, and was then added to the capital. Again, it is a useful exercise in scrutiny to ask students to try and explain the figures, as published, for the scheme. For such exercises to succeed, it is important that the students should be given consistent and appropriate advice on interest calculations throughout.

**2.3. Elephant traps**

Much of the emphasis in modern interactive design is on enabling users to work quickly by making choices, and entering data via structured input panels. This is fine, for users who know exactly what they are doing. For those who do not (i.e. by definition, most students) there are some dangers. The arrival of the **Insert Function** facility, for example, has led to various forms of 'function surfing". Students presume that there is a function for *everything.* It is generally easier to hunt around for a function to try, than to bring up the Help screen, (or even to keep your fingers off the mouse, and think it through for yourself). Having found what seems to be a suitable candidate, students will start to stuff all kinds of data into the panels of the wizard, in a determined (and usually increasingly bad-tempered) effort to persuade it to deliver the outcome they want.

There are two versions of the elephant trap. Firstly, the function may yield up a result which looks fine, but which is actually wrong. For example, **INTRATE** looks like a useful contender for finding an interest rate, given initial and final values for an investment, but only works for simple interest. Alternatively, the function may indeed be the right one, but if inappropriate data is fed into it, the rather restricted feedback may not alert the user to this. Indeed, it is all too easy to assume that the wizard is giving its approval simply by accepting the input. Unfortunately, quite simple errors may be overlooked. For example, it may be that 01/01/80 is being calculated as .125, rather than being interpreted as a date, or a percentage value is being taken at a hundred times the value intended. Of course students make exactly the same mistakes when entering values into cells. Their shift in perception is that the wizard is somehow all-knowing. They expect it to pick up errors for them.

### 3. IMPLICATIONS

All the instances mentioned above can be dismissed as being relatively trivial. But in financial calculations, quite minor misjudgements or misapprehensions on the part of the user can cause havoc in the final results. Consideration of these minor flaws also prompts a more worrying question. Even if we wish to correct some of the defects in existing functions, is it actually feasible to do so?

At this point, some sympathy is due to the software developers. Take the familiar Net Present Value function, for example. Ever since its inception (by Lotus, in this case) it has been a trap for the unwary. A text book published more than ten years ago advises: 1t is important to note that the Lotus @NPV routine applies discounting to *all* the cells in the specified range" [7]. This convention was followed by the designers of Excel. Any initial investment, which does not need discounting because it is *already* at today's values, should be omitted.

This does not seem at all intuitive to students, as can be seen by the number of them who take an entire series of cash flows, and feed them all into the NPV function. (They could of course use the more recent XNPV function, which discounts from the second payment onwards. However, this requires the user to enter both cash figures *and* a matching set of dates. Apart from the complexity of the function, they may also be puzzled by an enigmatic statement in the Help screen, to the effect that *"the first payment is optional')*.

It would be nice to re-invent the NPV function to be more intuitive, but this is a QWERTY-type problem. Experienced users expect NPV to work in a particular way, and it is buried in the calculations of countless existing spreadsheets. So we would already seem to be locked in to a particular way of doing things.

Similar issues arise in respect of conventions which vary between countries or between industries. A "UK Edition" of Excel might ensure that all calculations complied with UK practice. The Help screens could provide exact guidance on when to use particular ways of calculating interest or depreciation. However, such spreadsheets would not travel well.

Someone loading the spreadsheet into a US Edition would immediately have to be alerted to its original "nationality".

A final issue to be confronted is one which arises in a great deal of accounting software. Practices which have evolved in paper-based systems do not always make sense in electronic ones. Thus many accounting packages advertise themselves as "double entry", even though some of the error checking inherent in double entry methods is no longer done or required. Similarly, the survival of calculations based on 30 and 360 days, rather than true elapsed time, can no longer be defended on grounds of practicability. As Bhandari points out, what now seems to happen very often is that computer power is used to hoodwink the consumer, by using 360 as a divisor where it helps to top up the revenue of the loan issuer [8]. If combined with variable rates [9, 10]. The interest charges can become extremely difficult to audit. Logic suggests that everyone should use an identical "true" approach to interest periods, but this is an area where spreadsheets can only reflect, not lead, the real world.

**4. THE WAY AHEAD**

In the meantime, there are two strands of policy which perhaps should be considered. In education, it would help if more recognition were given to the necessity of spending time with students, discussing and explaining the nature of the tools they are using. This can only pay dividends in the way they apply them in their subsequent careers. Simply because spreadsheet technique can easily be learned from books and interactive tutorials, it does not follow that good practice follows close behind. At a time when Higher Education is constantly being urged to be more "efficient", it should be noted that cutting back on staff contact time, and relying substantially on automated teaching, will probably be a false economy in the longer term.

Secondly, it would help to have more constructive discussions between the designers of specialised spreadsheet facilities, such as the built-in financial functions, and those who set the standards in the relevant industry sectors. This implies that suppliers should look beyond merely checking that functions work exactly "as per specification" or adding in extra ones if there seems to be demand. We should be urging them to integrate the technical and the financial advice being given to users. This would mean calling on the expertise of outsiders such as accounting's professional bodies, and possibly developing different "dialects" for Help and user documentation for different countries. This would be a practical acknowledgement that errors are likely to follow not just from failures in technique, but from failures of alignment between the spreadsheet functions, and the conventions and perceptions of the world they serve.

REFERENCES

1. P. S. Brown, J D Gould, *An Experimental Study of People Creating Spreadsheets,* ACM Transactions in Office Information Systems, 1987, 5, 258-273.

2. R. R. Panko, *Hitting the wall. Errors in developing and code inspecting a "simple" spreadsheet, Decision Support Systems,* 1998, 22A, 337-353.

3. D. Janvrin, J Morrison, *Using a structured design approach to reduce risks in end user spreadsheet development,* Information & Management, 2000, 37A, 1-12.

4. F. Hormann, *Getting the oops! out of spreadsheets,* Journal of Accountancy, 1999, 188A, 79-83.

5. D. Whittaker, *Spreadsheet errors and techniques for finding them, Management* Accounting (UK), 1999, 77:9,50-51.

6. Microsoft, XL: *Adjusting the Period of Financial Functions,* ref Q36656, April 2000, www.microsoft.com.

7. M. Jackson, *Creative Modelling with Lotus 123,* Wiley, 1989, p. 194.

8. S. B. Bandhari, *Some ethical issues in computation and disclosure of interest rate and cost of credit, Journal of Business Ethics,* 1997, 16:5, 531-535.

9. J. Badger, *Adjusting to the Perils of ARMs,* Mortgage Banking, 1991, 51:11, 53.

10. J. Lankamp, *It pays to check lender errors,* Journal of Property Management, 1996, 61:2,